\documentclass[preprint2]{aastex}	

\newcommand{\av}{\mbox{$A_V$}}			
\newcommand{\gprime}{\mbox{$g^\prime$}}		
\newcommand{\hii}{\ion{H}{2} }			
\newcommand{\kms}{km~s$^{-1}$}			
\newcommand{\msun}{\mbox{${\cal M}_{\odot}$}}	
\newcommand{\n}{NGC~}				
\newcommand{\rprime}{\mbox{$r^\prime$}}		
\newcommand{\zo}{\mbox{$Z_{\odot}$}}		

\shorttitle{NGC~3256  Clusters}
\shortauthors{Trancho et al.}

\begin{document}

\title{Gemini Spectroscopic Survey of Young Star Clusters in
       Merging/Interacting Galaxies. II. NGC~3256  Clusters}  


\author{Gelys Trancho}
\affil{Universidad de La Laguna, Tenerife, Canary Island, Spain}
\affil{Gemini Observatory, 670 N. A'ohoku Place, Hilo, HI 96720, USA}
\email{gtrancho@gemini.edu}
\author{Nate Bastian}
\affil{Department of Physics and Astronomy, University College London,
       Gower Street, London, WC1E 6BT, United Kingdom}
\author{Bryan W. Miller}
\affil{Gemini Observatory, Casilla 603, La Serena, Chile}
\and

\author{Fran\c{c}ois  Schweizer}
\affil{Carnegie Observatories, 813 Santa Barbara Street, Pasadena,
       CA 91101-1292, USA}

\begin{abstract}

We present  Gemini optical spectroscopy of 23 young star
  clusters  in   \n3256. We find that the cluster ages range 
 are from few Myr to $\sim\,$150 Myr. All these clusters are relatively 
   massive  (2--40) $\times 10^{5}~\msun$ and appear to be of roughly 1.5~ \zo
  metallicity.  The majority of the clusters in our sample follow the same rotation curve 
as the gas and hence were presumably formed in the molecular-gas disk. However, a 
western subsample of five clusters has velocities that deviate significantly from the gas 
rotation curve. These clusters may either belong to the second spiral galaxy of the 
merger or may have formed in tidal-tail gas falling back into the system.  We discuss our observations in light of other known cluster populations in merging galaxies, and suggest that NGC~3256 is similar to Arp~220, and hence may become an Ultra-luminous Infrared Galaxy as the merger progresses and the star-formation rate increases.

Some of the clusters which appeared as isolated in our ground-based images are clearly 
resolved into multiple sub-components in the HST-ACS images. The same effect has 
been observed in the Antennae galaxies, showing that clusters are often not formed in 
isolation, but instead tend to form in larger groups or cluster complexes

\end{abstract}

\keywords{globular clusters: general ---
globular clusters: individual (\objectname{NGC 3256})}

\section{Introduction}

In the prevailing picture of hierarchical early-type-galaxy formation,
small fragments form first and then merge into larger and larger pieces
until the system resembles a large, smooth, anisotropically supported
elliptical galaxy.  The study of globular clusters has an important
role to play in testing the predictions of this theory and in
answering the question about if and when the bulk of this merging took
place.  Studies of the Galactic globular cluster system have been
fundamental to the development of ideas on how the Galactic halo, and
perhaps the entire galaxy, has been assembled from merging fragments
(Searle \& Zinn 1978; Da Costa \& Armandroff 1995).  Studies of
extragalactic cluster systems have revealed evidence for multiple
populations of clusters and allow us to use them to connect the
various phases of galaxy evolution.

GCs in galaxies more distant than about a megaparsec cannot be
resolved into stars, even with HST.  However, the fact that they
contain simple stellar populations means that their integrated
properties can be used to determine metallicities and ages.  
Surveys of large numbers of globular cluster systems reveals  that $>$90\% of all Es
  have bimodal color distributions, and only 10-20\%  in S0s (Kundu \& Whitmore (2001a, 2001b)).
 Almost all galaxies host a GC population with a peak in their color distribution at (V-I)=0.95.
Available metallicity measurements suggest that these are old,
metal-poor clusters like the Galactic halo GCs (Geisler, Lee, \& Kim
1996).  The second peak is redder than the first and implies that the
red GCs are more metal-rich.  In general, these GCs are also old, but
the age-metallicity degeneracy in broad-band colors makes age
differences hard to determine unless the GCs are rather young (see
Whitmore et al. 1997).

There are several competing scenarios for how the globular cluster
systems of galaxies formed.  \cite{ashan92} predicted that the
merger of two disk galaxies would produce an elliptical with a bimodal
GC color distribution.  The blue clusters are Galactic-halo like
clusters from the progenitor galaxies.  These clusters would have
formed like those in the Galaxy, perhaps from the accretion of dwarf
galaxies with large numbers of GCs.  Dwarf elliptical galaxies were
efficient at forming clusters and these clusters resemble those in the
Galactic halo (Miller et al. 1998).  In the Ashman \& Zepf scenario,
the red population is formed during the merger process from the
metal-enriched gas in the disks.  The formation of new clusters also
alleviates the problem that ellipticals have specific globular cluster
frequencies (number per unit luminosity) about a factor of two higher
than spirals do (Schweizer 1987).  An alternative scenario is that all
the clusters
were formed ``in situ,'' in a multi-phase collapse of a single potential
well (Forbes et al. 1997) similar to the early monolithic collapse
picture of the formation of the Galaxy (Eggen, Lynden-Bell, \& Sandage
1962).  In this case, the metal-poor clusters form first in the halo
and and metal-rich clusters form later during the final collapse, or
recollapse, stages from metal-enriched gas. Both these scenarios fit
into the overall picture of hierarchical galaxy formation, but they
differ in when the bulk of the merging takes place.

An important result of the merger scenario is that
the merger of two spiral galaxies can cause the formation of many
young star clusters in the starburst resulting from the collision of the
two gas disks (Schweizer 1987; Ashman \& Zepf 1992).
The creation of star clusters may alleviate the
problem that elliptical galaxies have specific globular cluster
frequencies (S$_N$, the luminosity-weighted number of GCs) about a
factor of two higher than in spirals.  The evidence is growing that
significant numbers of star clusters are formed during galaxy mergers.
Hubble Space Telescope observations of NGC1275 (Holtzman et al. 1992;
Carlson et al. 1998) were some of the first to find very luminous blue
objects with the properties of young GCs in a galaxy that may
have had a recent merger.  Large young cluster populations with ages
between 10 and 500 Myr were soon found in obvious merger remnants like
NGC7252 (Whitmore et al. 1993; Miller et al. 1997), NGC3921 (Schweizer
et al. 1996), NGC3256 \citep{zepf99}, NGC4038/4039 (Whitmore \&
Schweizer 1995; Whitmore et al. 1999).  Spectra of the brightest young
clusters in NGC1275 (Zepf et al. 1995) and NGC7252 (Schweizer \&
Seitzer 1993,1998) confirmed that the clusters were between 0.5 and 1
Gyr old with roughly solar metallicities (Figure 2).  Since these
clusters are many internal crossing times old, they would seem likely to
survive to become the red GC populations of the faded merger remnants
(e.g., Whitmore et al.\ 1997; Goudfrooij et al.\ 2001a).

These observations allow us to proceed to the next level of
understanding the evolution of globular cluster systems; we must
understand in more detail how star clusters form and how systems of
star clusters evolve.  It is thought that star clusters are formed in
high-pressure environments (Harris \& Pudritz 1994; Elmegreen \& Efrimov
1997), be they caused by the collisions of giant molecular clouds
or by a general pressure increase in the gas surrounding molecular clouds.
Observations of
the Antennae can now provide some key parameters about the state of
the ISM during clusters formation and the feedback produced by the
young clusters.  Zhang, Fall, \& Whitmore (2001) compared the
locations of different cluster populations with emission from the ISM
at wavelength from X-rays to radio and found that the youngest
clusters are associated with molecular cloud complexes and may lie in regions
of high HI velocity dispersion.  However, the small velocity dispersion
of the clusters among themselves strongly suggest that it is not
high-velocity cloud-cloud collisions that drive cluster formation, but
the general pressure increase experienced by gas during the merger
(Whitmore et al.\ 2005).
Feedback, seen in the form of H$\alpha$ bubbles around young-cluster
complexes, may enhance this process.

An interesting question is how globular cluster systems evolve.
Most very young star-cluster systems studied in merging galaxies have
power-law luminosity functions of the form M$^{-2}$ (e.g. Schweizer et
al. 1996; Miller et al. 1997; Whitmore et al. 1999; Zhang \& Fall
1999).  However, the luminosity (or mass) function of old globular
clusters has a log-normal shape with a peak at $M_V^0 \sim -7.3$
($\sim10^5$ M$_{\odot}$; Harris 1991).  Therefore, either the initial
mass function of star clusters was different in the past, perhaps due
to low metallicity, or a substantial fraction of the young clusters
must be destroyed for the young mass functions to evolve into what we
see in older systems.  A great deal of theoretical work has gone into
globular cluster destruction processes (e.g. Fall \& Rees 1977; Gnedin
\& Ostriker 1997; Vesperini 1997/98; Fall \& Zhang 2001).  The main
processes that can destroy globular clusters are stellar evolution,
2-body relaxation, dynamical friction, and disk shocking,  or if the clusters
 formed in gas rich environments, interactions with GMCs can play 
 a significant role (Gieles et al.~2006).  The models
of Fall \& Zhang (2001) predict that there may be radial variations in
the peak of the mass function within a galaxy and that the peak will
shift to higher masses with time.  Many of these processes depend on
the relative velocities of the clusters and field stars and on the
cluster orbits.

In this paper we present some first results of a large spectroscopic survey of
star clusters in merging and interacting galaxies.  We focus on 23
bright star-cluster candidates in the main body of \n3256 observed
with GMOS on Gemini South.  Other targets in our survey, to be
presented in future papers, include NGC~4038/39 (The Antennae),
\n6872, Stephan's Quintet, and M82.

\n3256 was classified as an intermediate-stage merger in Toomre's list
of nearby merging systems \citep{toomre77}.  The merger is more
advanced than systems like NGC 4038/39, in which the two original disks
are still distinct, but the two nuclei (Moorwood \& Oliva 1994;
Norris \& Forbes 1995; Lira et al.\ 2002; Neff et al.\ 2004) have not
merged yet either.  They have a separation in projection by 5", or
$\sim900$~pc.  Hence, \n3256 is not as relaxed a system as \n7252 is.
The outer parts of the system are characterized by shell-like features
and two
extended tidal tails that are typical of merging galaxies.  The body
of the system is criss-crossed by dust lanes that enshroud an on-going
starburst: the far-infared luminosity, X-ray luminosity, and star
formation rate are the highest of all the systems in the Toomre
sequence.  This starburst has created over 1000 star clusters in the
central region \citep{zepf99} as well as in the tidal tails
\citep{knier03, trancho07a}.

  The current paper is organized as follows: Section~\ref{sec:obs}
describes the observations.  In \S~\ref{sec:results} and
\S~\ref{sec:kinematics} we derive the ages, masses, extinctions,
metallicities, and line-of-sight velocities of the 23 clusters.
Finally, \S~\ref{sec:disc} discusses and summarizes the results.

\n3256 is located at $\alpha_{\rm J2000}=10^{\rm h}27^{\rm m}51\fs3$,
$\delta_{{\rm J}2000}=-43\degr54\arcmin14\arcsec$ and has a recession
velocity relative to the Local Group of $cz_{_{\rm Helio}} = +2804\pm 6$
\kms\ , which places it at a distance of
36.1 Mpc for $H_0 = 70$ \kms\ Mpc$^{-1}$.
At that distance, adopted throughout the present paper, $1\arcsec = 175$ pc.
The corresponding true distance modulus is $(m-M)_0 = 32.79$.
Because of the low galactic latitude of \n3256, $b = +11\fdg7$, the
Milky Way foreground extinction is relatively high, $A_V=0.403$ (Schlegel
et al.~1998), whence the apparent visual distance modulus is $V-M_V = 33.19$.

\section{Observations and Reductions}
\label{sec:obs}

Imaging and spectroscopic observations of star clusters in \n3256 were
made with GMOS-S in semesters 2003A and 2004A.
The data were obtained as part of two Director's Discretionary Time
programs, GS-2003A-DD-1 and GS-2004A-DD-3.
Our images cover the typical GMOS-S field, which measures approximately
$5\farcm5\times  5\farcm5$.
They were obtained through the \gprime\ and \rprime\ filters.
Four GMOS masks with slitlets were used for the spectroscopic observations.
We used the B600 grating and a slitlet width of $0\farcs75$, resulting in
an instrumental resolution of 110 km/s at 5100 \AA.
The spectroscopic observations were obtained as 8 individual exposures
with exposure times of 3600 sec each.
Our spectroscopy of 70 cluster candidates yielded only 26 objects that
were bonafied star clusters in \n3256.
Of these, three are located in the western tidal tail and have been
described in Trancho et al.\ (2007).
In the present paper we focus on the 23 star clusters located in or near
the main body of NGC~3256.
Table~\ref{table:properties1} lists these star clusters.
Column (1) gives the adopted cluster ID, columns (3)--(4) the coordinates,
and columns (5) and (6) the absolute magnitudes $M_{g'}$ and $M_{r'}$ and
their errors.
The magnitudes have been corrected for Galactic extinction, but not for
any internal extinction.

Figure~\ref{fig:image} shows an {\em HST}/ACS image of the main body of
\n3256, with the observed candidate clusters marked by their ID numbers. 

The basic reductions of the data were done using a combination of the
Gemini IRAF package and custom reduction techniques, as described in
Appendix A.

\section{Derivation of Cluster Properties}
\label{sec:results}

The derivation of cluster properties (such as age and metallicity) based on the strengths of 
stellar absorption lines through optical spectroscopy  is not straightforward, due to 
degeneracies between age, metallicity, and extinction.
Multiple studies have addressed this problem (e.g., Schweizer \& Seitzer
1998; Schweizer, Seitzer, \& Brodie 2004; Puzia et al.\ 2005), and here we extend previous studies.

Although in some of our cluster spectra the strengths of stellar absorption lines cannot be measured due to strong emission lines, the
fluxes/equivalents width(EW) of the emission lines of the surrounding \hii region can be
measured. In these cases a chemical abundance can be estimated for the \hii
region, in which the cluster has recently formed, from line-emission measurements (e.g., Kobulnicky \& Kewley
 2004; Kobulnicky \& Phillips 2003;  Vacca \& Conti 1992). 
  This abundance is expected to be the same as that of the young stellar cluster itself,  and
hence complements abundance measurements of the absorption-line
dominated clusters.  

Below, we outline the method adopted in the present study.

\subsection{Extinction, Age, and Metallicity}
\label{sec:ages}

\subsubsection{Absorption-Line Clusters}
We first select the model spectra by Bruzual \&
Charlot (2003, hereafter BC03) and by  Gonz\'alez-Delgado et al.\ (2005,
hereafter GD05) . Then we smooth the model spectra to the same resolution
as the observed spectra. Then we compare the cluster spectra with the models
for clusters of solar metallicity, to which we have applied
various amounts of extinction (using de Galactic extinction law for Savage \& Mathis 1979) , $A_V = 0$--10 in steps of 0.1 mag.
We select the best fitting model via minimized $\chi^2$,
Model$_{\rm best}$(age,\av), and correct the observed spectrum for the
derived extinction $\av$ to yield Cluster$_{\rm obs,ext}$.

This procedure was carried out for the BC03 and GD05 models independently,
and we note that for individual clusters the results are very similar.
Due to the finer grid of young ages in the GD05 cluster models we adopted
these for our further analysis.

We then inserted the extinction-corrected spectra into the IDL
implementation of the Penalized Pixel Fitting routine (pPxF) of
Cappellari \& Emsellem(2004)\footnote{We used the GD05 spectra which have a resolution of 0.3 \AA ~at 5100\AA
, corresponds to $\sigma \sim 0.3/5100*3e5/2.35 = 7.5$ km/s and 
our spectra have a   $\sigma \sim 3.96/5100*3e5/2.35 = 99.12$ km/s.
The quadratic difference is sigma necessary to use in ppxf is  $\sigma =\sqrt(99.12^2-7.5^2)= $ 98.83 km/s
}.
This routine determines the best linear combination of template spectra
plus an analytic polynomial to reproduce the observed spectra and returns,
in addition, the radial velocity of each cluster.
For template spectra we used all available models (both in age and
metallicity) of GD05.
We suppressed the use of any additional polynomial in order to preserve the continuum
shape of each observed cluster.
Emission lines and regions of the spectra affected by artifacts were masked
during the fits, though care was taken to minimize the number of such areas.
Via this procedure we obtained a template spectrum for each of our clusters,
Cluster$_{\rm temp}$, as well as an emission spectrum (the difference between
Cluster$_{\rm obs,ext}$ and Cluster$_{\rm temp}$).

In order to find the age and metallicity of each cluster, we measured
the line strengths of the hydrogen Balmer lines as well as of prominent
metal lines (See Table 2).
For this we used the Lick indices (Faber et al.\ 1985; Gonz\'alez 1993;
Trager et al.\ 1998) as well as the indices defined by Schweizer \& Seitzer
(1998) for young stellar populations.
To make the measurements, we used the routine {\em Indexf} (Cardiel et
al.\ 1998), which finds the line strengths and errors by performing
Monte-Carlo simulations on the spectra, using information derived from
the error spectra (i.e., the placement of the continuum bands and noise
in the data).
{\em Indexf} was run on Cluster$_{\rm temp}$ instead of
Cluster$_{\rm obs,ext}$.
The reason for this is that we found that the measured index strengths of
the Cluster$_{\rm obs,ext}$ depended on the S/N ratio of the spectrum.
This is shown in Fig.~\ref{fig:test-ppxf-obs}, where we plot the nine
measured indices of an observed cluster (T130 in The Antennae, which is our best S/N cluster and
 has identical setup and was observed in the same way) degraded to various S/N ratios. 
The open symbols represent the measurement of the line strength for each
index when {\em Indexf} is run on Cluster$_{\rm obs,ext}$, while the filled
symbols represent the measurements carried out on Cluster$_{\rm temp}$.
The lines show the average index strength for the five highest S/N
measurements.
From this numerical experiment we conclude that the measurements on
Cluster$_{\rm temp}$ reproduce those of Cluster$_{\rm obs,ext}$ for
high S/N, but remain accurate to S/N $<$10, while measurements on
Cluster$_{\rm obs,ext}$ begin to show significant scatter below
S/N $\approx$ 15.

However, we note that with this adopted procedure the measured H$\beta$
index is always systematically off.
The models never reproduce an absorption
feature on the red side of the line.
Additionally, similar tests were performed where we inserted the model
spectrum (i.e. treating the models as observations, degrading
them in S/N and finding the indices).  These tests showed a systematic offset
 in the H$\delta_A$ measurements although the other lines were well reproduced.
Therefore, we removed both the H$\beta$ and H$\delta_A$ indices from
our list for further analysis.

Using {\em Indexf}\,\ we also measured the line strengths of the GD05 models,
for all model ages and metallicities.
The age and metallicity of each observed cluster was then determined
by comparing its line indices to that of models, weighted by the respective 
errors, in a least $\chi^2$ sense.
In order to test the robustness of this technique we added random errors
to the measured indices (using a normal distribution with a dispersion
corresponding to the 1-$\sigma$ measurement error) and re-did the analysis.
This was done 5000 times for each cluster.
The final age was then determined by creating a histogram of the derived
ages and fitting a gaussian to it (in logarithmic space), where the adopted
age is the peak age of the gaussian and its error is the standard deviation.
The cluster metallicity was found by simply averaging the derived
metallicity of the 5000 runs.
Examples of this process are shown in Fig.~\ref{fig:ages}.

Finally, to check the consistency of our results, we plotted the spectra
of Cluster$_{\rm obs, ext}$ and the best fitting  model (i.e., the
model closest in age and metallicity).
If the fit was not satisfactory, then we began the entire process again,
eliminating the initially derived extinction from the options.
This was the case for only a handful of young clusters that were initially
fit with low extinctions and higher ages.
Two examples of cluster spectra are shown in Fig.~\ref{fig:examples}.

The black lines are the Cluster$_{\rm obs, ext}$ spectra, while the red
and green lines represent the best fitting model (age, metallicity) and
the residual (Cluster$_{\rm obs, ext}$ $-$ template $-$ constant).
T1002 in the right panel is clearly very young and has, as such, still
ionized gas associated with it.
In the observed and residual spectra, we clearly see emission lines from
H$\gamma$, H$\beta$ and [OIII]$\lambda\lambda$4959,5007.

As a further test of the derived cluster ages, we chose a subset of
indices that are good at distinguishing between the age and metallicity
of a cluster.
For illustrative purposes, Fig.~\ref{fig:indices} shows the [MgFe] index
(Thomas et al. 2003) plotted versus the H$\gamma$ index for both the GD05 and BC03
models.   The indices from the models are also shown. From these
  diagrams we check for consistency in the derived ages and
  metallicities of the clusters.

The derived ages, extinctions, and metallicities are given in
Table~\ref{table:properties1}.

\subsubsection{Emission-Line Clusters}

For the youngest clusters with little or no absorption features in their
spectra the task is much easier.
First, we assign ages to these clusters of less than 10~Myr, due to the
presence of large amounts of ionized gas around the cluster.
Age dating can be refined to some degree by the presence or absence of
Wolf-Rayet features (e.g., Cluster T2005, see Fig~\ref{fig:wr}); however,
that is beyond the scope of the present paper.
The extinction of these cluster is calculated from the H$\gamma$ to H$\beta$
emission-line ratio.

We adopt the chemical analysis method from Kobulnicky \& Kewley 2004 (hereafter KK04) to determine 
the metallicity.
We measure the EW ratio of the collisionally excited 
[OII]$\lambda$3727 and [OIII]$\lambda\lambda$4959,5007 emission lines relative to the H$\beta$
recombination line (known as R$_{23}$) and [OIII]$\lambda\lambda$4959,5007  relative to  [OII]$\lambda$3727 
 (knows as O$_{32}$) , along with the calibrations on KK04 (their Fig. 7 - upper brach) and
the solar abundances by Edmunds \& Pagel (1984).
Instead of the traditional flux ratio, the KK04 method uses EW ratios that have the
  advantage of being reddening independent.

As can be seen in Table~\ref{table:properties1} the metallicites found
for absorption-line and emission-line clusters agree well, giving us
confidence in the robustness of the diagnostic methods and results. 

\subsection{Masses}

In order to calculate the mass of each cluster we compared the photometry
(\gprime\ and \rprime) with the BC03 SSP models, assuming a Chabrier (2003)
stellar initial mass function and solar metallicity.
We then used the age dependent mass-to-light ratio from the models to
convert our derived absolute magnitudes (observed magnitudes corrected for
Galactic extinction, internal extinction, and the assumed distance modulus)
to mass.
Errors on the mass were estimated from the derived errors on the age and
photometric errors.
Systematic errors (e.g., errors associated with the distance to \n3256) are
not included.
Table~\ref{table:properties2} gives the derived masses of the clusters.

\subsection{Velocities}

In the case of absorption-line clusters, we used the IRAF task
{\em rvsao.xcsao} for the determination of the redshift from the
individual spectra, using three different type (A, O, B) radial-velocity standard stars (HD~100953,
HD~126248, and HD~133955) observed at the same resolution as the clusters.
The three template stars were employed to reduce the
systematic errors introduced by the effect of template mismatch when
computing the redshift using the cross-correlation technique.

For the emission-line clusters, velocities were measured from the observed
emission lines using the IRAF task {\em rvsao.emsao}.

In both cases the velocities were corrected to heliocentric (see Table 3).

Figure~\ref{fig:positions} shows the positions of the observed clusters
within \n3256 (shown in contours to highlight its main features), marked
with the cluster metallicities, extinctions, ages, and velocities,
respectively.

\section{Cluster Kinematics:  Two Populations?}
\label{sec:kinematics}

There is strong evidence that the molecular gas in the central region
of \n3256 lies in a disk that rotates (Sakamoto et al.\ 2006).
The rotation axis of this gas disk lies approximately along the
north-south direction, which is also the apparent minor axis of the main
optical disk.
It is interesting to compare the observed cluster radial velocities to
the molecular-gas velocities at each cluster's position.

Figure~\ref{fig:rot-vel} shows the measured radial velocities of the clusters
plotted versus the cluster right ascension (RA), corresponding roughly
to their projected position along the major axis.
Superimposed is the rotation curve for the molecular gas measured by
Sakamoto et al.\ (2006).
The figure suggests that the majority of the clusters are still associated
with the gas from which they formed (see Table~\ref{table:properties2}).
At least 15 of the 23 observed clusters show clear evidence of corotating
with the molecular-gas disk.  Hence, we will refer to these clusters as ``disk clusters.''

\subsection{Origin of the Disk}

Sakamoto et al.\ (2006) suggest that the molecular-gas disk may have
formed during the merger of the parental spiral galaxies. 
 By using the ages of the disk clusters we can put a lower limit on the
longevity of the disk.  These ages range from recently formed (e.g., T761: $<$10~Myr) to 100~Myr
old (T112).  Thus the molecular-gas disk must have existed for at least 100~Myr.

The \n3256 merger probably began approximately $\sim500$ Myr ago (English et al 2003).
It is not yet completed, as the two nuclei have yet to merge (Moorwood \&
Oliva 1994; Norris \& Forbes 1995; English \& Freeman 2003).
Therefore, {\em if}\, the observed molecular-gas disk and clusters formed
during the merger, the disk must have begun forming early in the merger.
Sakamoto et al.\ (2006) compare the \n3256 system to that of \n7252, a
recently formed merger remnant which also hosts a molecular-gas disk.
However, in contrast to the situation in \n7252 the two nuclei of \n3256
have yet to merge, which may disrupt any current large-scale gaseous
disk (Barnes 2002). 
Hence, the two merger systems may not presently be comparable.
An alternative hypothesis, however, is that the observed gas disk belongs
to one of the two original spiral galaxies, so that the observed disk
clusters simply formed in that disk as part of the enhanced star-formation
activity caused by the gravitational interaction.  

In either scenario, we would expect older star clusters to be present as
well.  The fact that they are not detected in the present study is readily
explained by selection bias: we selected the brightest clusters, which
tend to be young, for spectroscopy.

\subsection{Non-Disk Clusters}

In the western section of the galaxy we find five clusters which have
velocities apparently inconsistent with an extrapolation of the
molecular-gas disk velocities.  The clusters have ages 
 between $<7$~Myr (e.g. T96) and $\sim150$~Myr (e.g. T1078).  These clusters may belong
either to the other spiral disk (which may lie behind the observed disk
in projection) or to material which has become dissociated from the
original disks due to the interaction/merger.  These clusters are also located spatially near the
 beginning of the western tail  (see Fig.~1 in Paper~I).
A detailed comparison with the HI position-velocity diagram of English et
al.~(2003) shows that the HI tail begins approximately $45\arcsec$ to the
west of the observed clusters.  However, as Fig.~\ref{fig:english-himap}
shows, these clusters are coincident spatially and kinematically with
HI gas that has a very different radial-velocity distribution from that of
the molecular gas.  The H I radial velocities reach a minimum near the
CO rotating disk and a maximum at the kinematic center of \n3256, whereupon
they begin dropping again inside the western tidal tail.
One possible interpretation is that the gas-velocity anomaly, noted already
by English et al.\ (2003), is caused by gas falling back into the central
parts of \n3256 from one of the tidal tails (perhaps the eastern one).

Two other clusters located closer to the observed galactic center stand out in terms of their 
kinematics.  These clusters (T779 and T343) have velocities larger than that expected 
if they were part of the rotating molecular-gas disk (although T343 is only incompatible with the disk velocity at the
 1.5$\sigma$ level).  Both clusters are very young and have only modest extinction (\av = 0.4--0.8).   
 We note that these two clusters are located in a part of the galaxy where there is a rather large 
 scatter in the measured velocities of the clusters (e.g. T201, T343, T356, T779) and thus their 
 deviation may be part of a larger trend.  It is possible that we are seeing a heating or beginning destruction of the disk
 due to the interaction/merger, and that star-formation is proceeding from an ordered phase, i.e. 
 in a disk, to that of a more chaotic phase where dispersion dominates over rotation.

\section{Discussion}
\label{sec:disc}

\subsection{Environment of the Clusters}
Some of the emission line clusters that appeared isolated in our ground based images
are clearly resolved into multiple subcomponents in the {\em HST}/ACS
images.
The same phenomenon has been observed in the Antennae galaxies (e.g.,
Whitmore \& Schweizer 1995; Bastian et al.~2006), showing that clusters
are often not formed in isolation but instead tend to form in larger
groupings, or cluster complexes.
These complexes are thought to be a short-lived phenomenon as they
disperse on short timescales, although merging within the central parts
of the groupings is possible (e.g., Fellhauer \& Kroupa 2002).
The remnants of such cluster-cluster mergers are an attractive means to
form extremely large clusters, such as W3 and W30 in NGC~7252
(Kissler-Patig et al.\ 2006).

Hence, it is possible that in a few cases (which are in very crowded
regions, e.g. T2005 and T116)  we may be over-estimating the mass of an apparent ``cluster''
if, in fact, it is made up of several clusters.

\subsection{Star/Cluster Formation Rates}

\n3256 contains the most molecular gas among the merging galaxies and merger
remnants of the Toomre sequence ($1.5\times 10^{10}$~\msun: Casoli et al.\
1991; Aalto et al.\ 1991; Mirabel et al.\ 1990, from Zepf et al.\ 1999).
Thus, there will be plenty of gas left to fuel a massive starburst when the
nuclei merge (e.g., Mihos \& Hernquist 1996).
At that time, one may expect the star/cluster formation rate to increase
substantially (between 3 and 10 times depending on the encounter parameters
and the time to nuclear coalescence) (Cox et al.~2006).
The present star-formation rate in \n3256 is estimated to be 33 \msun/yr
from the far-infrared luminosity (Knierman et al.\ 2003).
It is known that a tight relation between the most massive star cluster
in a galaxy and the galaxy's star-formation rate exists (e.g., Larsen 2004).
In \n3256 we find about 10 clusters with masses in excess of $10^6 M_{\odot}$.
If the star (cluster) formation rate does increase substantially as the
nuclei merge, we may expect the \n3256 system to create clusters
with masses significantly above 10$^7$\,\msun, such as those found in
\n7252 and \n1316 (e.g., Schweizer \& Seitzer 1998; Maraston et al.\ 2004;
Bastian et al.\ 2006).

Arp 220 is comparable to NGC~3256, as it also has an extremely high infrared luminosity, 
is thought to have formed through a merger, and has two distinct nuclei separated by only 300~pc in projection
(Scoville et al.~1998, Wilson et al. 2006).  The nuclei in NGC~3256 are separated (in projection) 
by $\sim900$~pc indicating that it is possibly slightly younger (in terms of merger stage) than Arp 220.
The star-formation rate in Arp~220 ($240~M_{\odot}$/yr; Wilson et al.~2006)
is approximately seven times higher than that in NGC~3256.  It is then
possible that NGC~3256 is currently 
poised to enter the regime of ultra-luminous infrared galaxies (ULIRGs) as its star-formation 
rate increases due to the merging of the two nuclei.  Arp 220 also closely follows the relation 
between global star-formation rate in the galaxy and the magnitude of the most massive cluster
 (Wilson et al.~2006), arguing further that NGC~3256 is going to form clusters in excess of $10^7 M_{\odot}$.

\subsection{Metallicities}

As Figure~\ref{fig:hist} shows the young clusters in the NGC~3256 system appear to
 have rather high metallicities, with the average being $\sim1.5 \zo$.  
 This was also found for the clusters in the tidal tails described in Paper~I.
We do not see any major spread in metallicities among our clusters and,
specifically, no differences coming from ages or emission 
versus absorption.  The fact that the majority of these young clusters formed in a 
disk and their age spread is small fits in nicely with the notion of a normal starbust process.

\section{Comparison of NGC~3256 with Other Merging Galaxies}

Many of the clusters observed to be associated with the molecular gas
are quite massive ($>10^5 \msun$), have survived for many internal
crossing times ($t_{\rm cr} \approx 2$--4 Myr), and are therefore
gravitationally bound.  
This justifies calling them young globular clusters.  Such young globulars
have been found in many merging galaxies, from beginning mergers (e.g.,
\n4038/39: Whitmore \& Schweizer 1995) to completed mergers (e.g., \n1316:
Goudfrooij et al.\ 2001a).   The formation of these clusters is 
thought to trace the major star-formation events in these galaxies and they must form with approximately 
the same kinematics as the gas out of which they form.  It is therefore interesting to compare the \n3256 
cluster population to those of younger (in terms of dynamical stage) and older merging systems.

The majority of the cluster population of the beginning merger \n4038/39
is still unambiguously confined to the disks of the two progenitor galaxies
(Whitmore et al.~2005; Bastian et al.~2006; Trancho et al.~2007).
Therefore, \n3256 appears to predominately fall into this category (see
\S~\ref{sec:kinematics}).

Older systems, on the other hand, such as \n3921 (Schweizer, Seitzer, \&
Brodie~2004), \n7252 (Schweizer \& Seitzer 1998), and \n1316 (Goudfrooij
et al.~2001b) are all characterized by the kinematics of their clusters
being dominated by non-circular, halo-type orbits.
When does the transition happen?  It will be interesting to determine
whether the majority of the star formation happens in the disks of the
progenitors and their orbits are subsequently randomized (i.e., turned
into pressure supported systems rather than rotational supported systems),
or whether the star-formation events which precede the destruction of the 
galactic disks pale in comparison with the star-formation rate during and
after the destruction.

Note that shortly after the merger a gaseous disk can reform around the
nucleus of the merger remnant, which can harbor subsequent star formation
(e.g., \n7252: Miller et al.\ 1997;  Wang, Schweizer, \& Scoville 1992),
although such star formation appears to occur at a much lower intensity
than previous star-forming episodes during the merger.

\section{Summary and Conclusions}

We have studied the ages, metallicities, masses, extinctions, and velocities
of 23 clusters in \n3256 based on the Lick index system in conjunction
with CO and HI maps.
The main results are:
\begin{itemize}
\item  The clusters have rather high metallicities, with the average being  $\sim1.5$\zo\ (Fig.\ 9) and are massive, with masses in the range (2--40) $\times 10^{5}~\msun$.  
The ages of the clusters are between a few Myr and $\sim\,$150 Myr.

\item  There is strong evidence for a rotating molecular-gas disk in
\n3256 (Sakamoto et al.~2006). 
The majority of the clusters in our sample follow the same rotation curve
as the gas and hence were presumably formed in the molecular-gas disk.
However, a western subsample of five clusters has velocities that deviate
significantly from the gas rotation curve.  These clusters may either
belong to the second spiral galaxy of the merger or may have formed in
tidal-tail gas falling back into the system.

\item Although the merger began $\sim\,$500 Myr ago (English et al.\ 2003),
we found the clusters to be $\lesssim150$ Myr old.  Since there are still
two distinct nuclei marking the presence of two galaxies, we conclude that
the gas disk probably belongs to one of the galaxies and is not yet a disk
of pooled gas
produced in the merger itself.  Presumably clusters older than the ones
present in our sample do exist in \n3256.  However, these older clusters
would not have been selected as spectroscopic candidates due their fainter
magnitudes (i.e., only the brightest candidates were selected).

\item  By  comparing of the NGC~3256 cluster population with other known galactic mergers, we suggest that this system is akin to Arp~220, although slightly dynamically younger.  If this is the case, the we may expect the star/cluster formation rate to increase significantly as the two galactic nuclei merge.  This in turn may push \n3256 into the category of ULIRGs (it is currently a LIRG).  Due to the expected large increase the in the star/cluster formation rate, a few clusters above $10^7$~\msun are predicted to form before this merger is through.

\item Some of the clusters which appeared as isolated in our ground-based
images are clearly resolved into multiple sub-components in the HST-ACS
images.   The same effect has been observed in the Antennae galaxies,
showing that clusters are often not formed in isolation, but instead tend
to form in larger groups or cluster complexes. 

With these new results, i.e.~cluster ages, metallicities, extinctions and kinematics, as well as recent CO  and HI maps, N-body simulations of this merger would be the best way to fully understand this wealth of data.  The models would have important implications for (globular) cluster formation and destruction as well as the star-formation history of the merger (through the age/metallicity of the clusters) with respect to other mergers like the Antennae, \n7252 and Arp 220.  Finally, the details of the models may present important implications of the formation of ellipticals galaxies through the major mergers of spiral galaxies.

\end{itemize}


\begin{figure}
 \begin{center}
   \epsscale{1.0}
       \caption{ ACS  F555W image of \n3256  (central region) with the
       observed candidate cluster ID numbers.  Green labels denote clusters with spectra dominated by emission lines,
        while red labels denote absorption line dominated cluster spectra.  The line with an arrow points north, while the line
without one points east.}
   \label{fig:image}
   \end{center}          
   \end{figure}

\begin{figure}
     \epsscale{1.1}
	\plotone{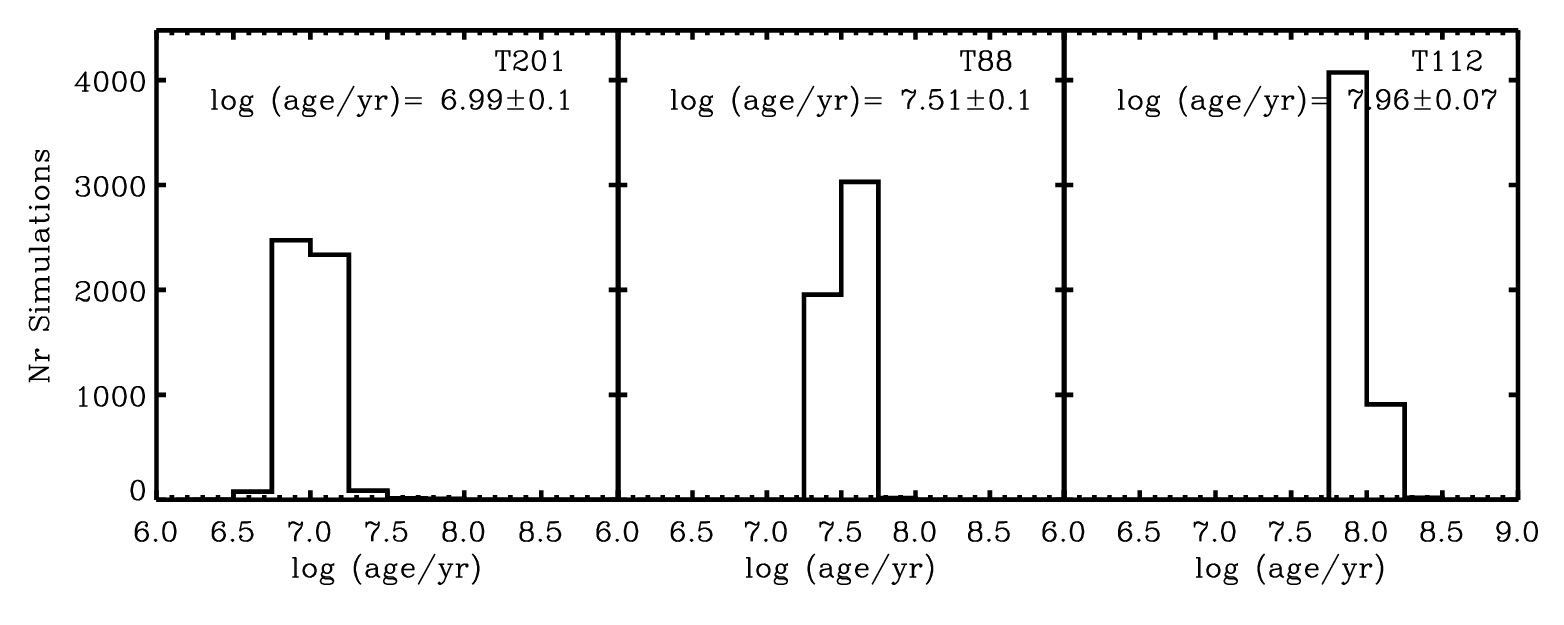}
	\plotone{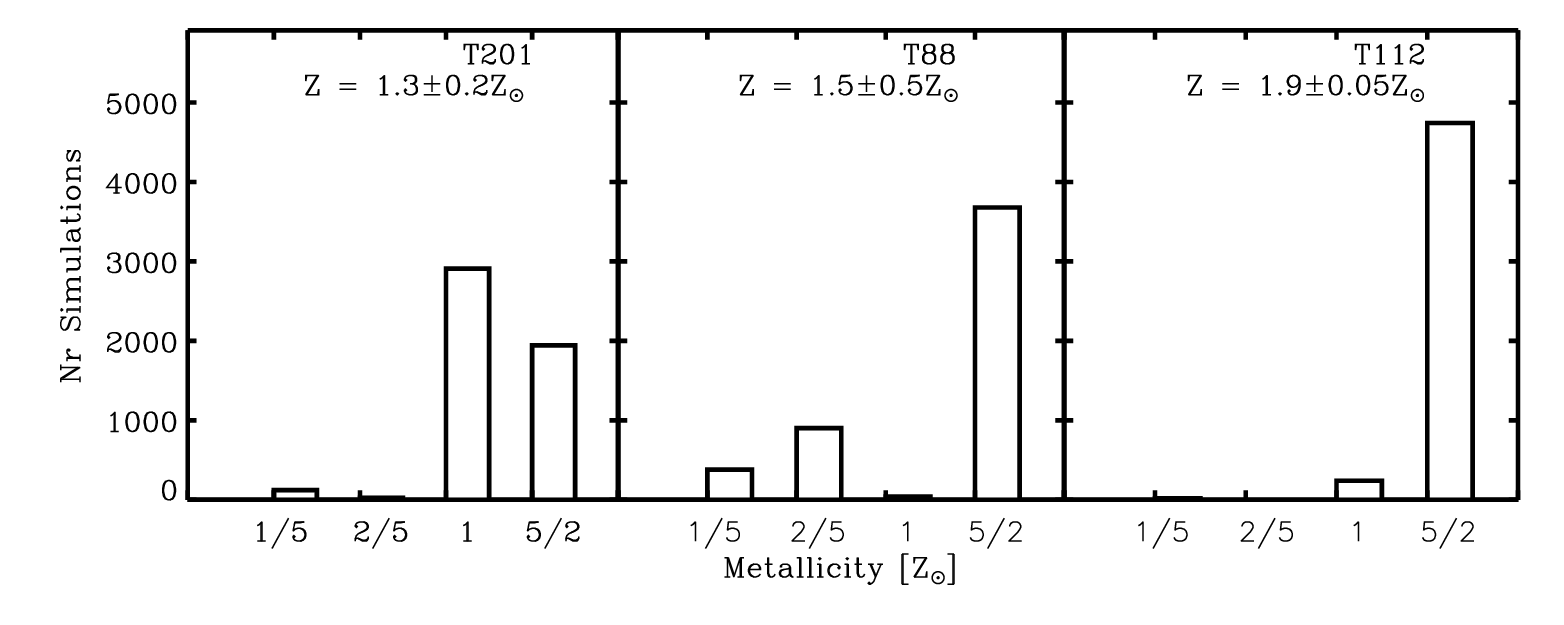}
    \caption{Top: Histograms of ages derived by simulating the
     effect of errors on the age-fitting routine.   The derived age and
     error (in logarithmic units) is given in each panel.   Bottom: 
     Same as top, except now for metallicity.  See text for details of
     the simulations.}
   \label{fig:ages}
   \end{figure}

\begin{figure}
   \epsscale{1.0}
     \plotone{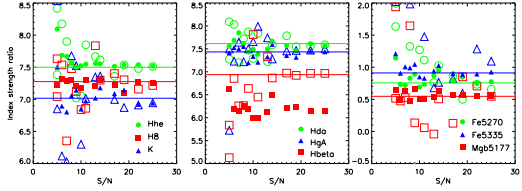}
      \caption{Tests showing the effect of the S/N ratio on the measured
       line indices.  We used a high-S/N cluster spectrum, degraded its S/N
       ratio, and then measured line indices.  Open symbols mark measurements
       made from the observed spectrum (corrected for extinction) directly,
       while solid symbols represent measurements made from a template
       spectrum derived for the cluster using the pPxF technique.  Solid
       lines represent averages of the highest S/N experiments on the
       observed cluster spectrum (Cluster$_{\rm obs,ext}$).  For further
       details, see text.}
   \label{fig:test-ppxf-obs}
   \end{figure}

\begin{figure}
     \epsscale{1.10}
  \plottwo{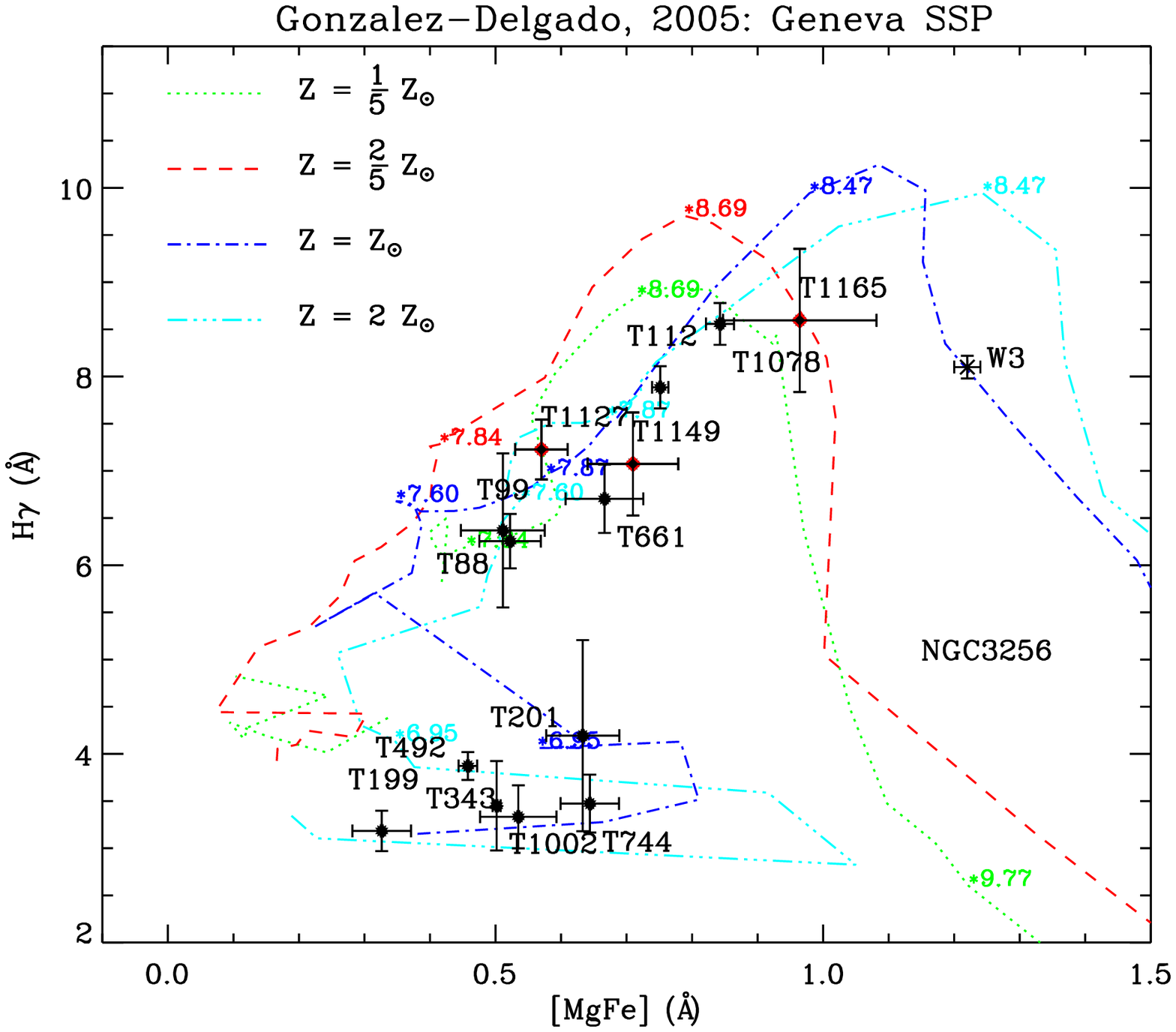}{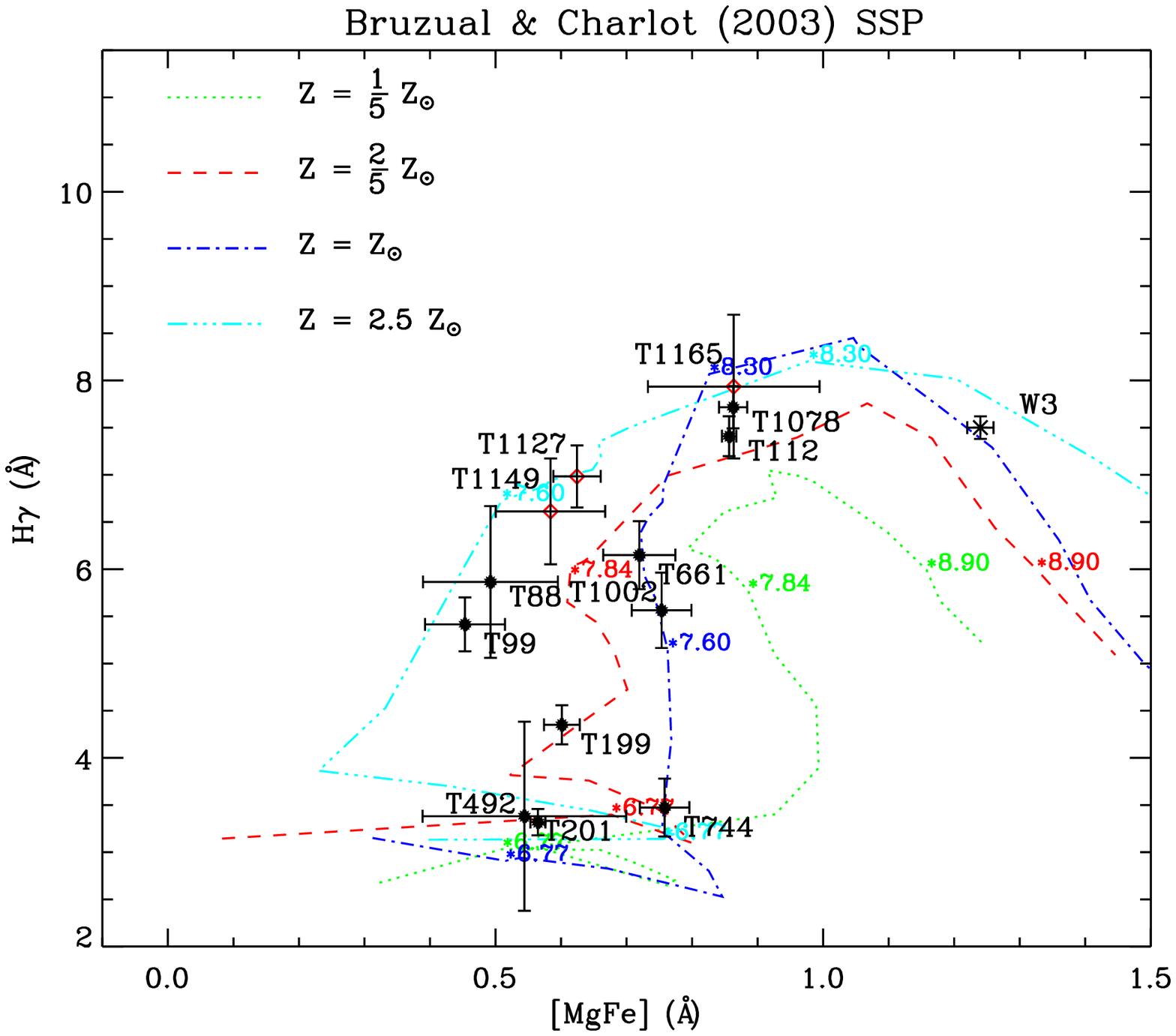}
      \caption{Determination of cluster ages and metallicities.
      (left) H$\gamma$ vs. [MgFe] from the Gonz\'alez-Delgado et al.\
      (2005) SSP models for four different metallicities are shown.
      Data points with error bars mark observed clusters and their
      1-$\sigma$ errors.  In addition, we show the position of the
      massive cluster W3 in NGC~7252 and the three tidal tail cluster in NGC~3256 from previous studies.
      (right) Same, but for Bruzual-Charlot (2003) models.}
   \label{fig:indices}
   \end{figure}

 \begin{figure}
   \epsscale{1.10}
    \plottwo{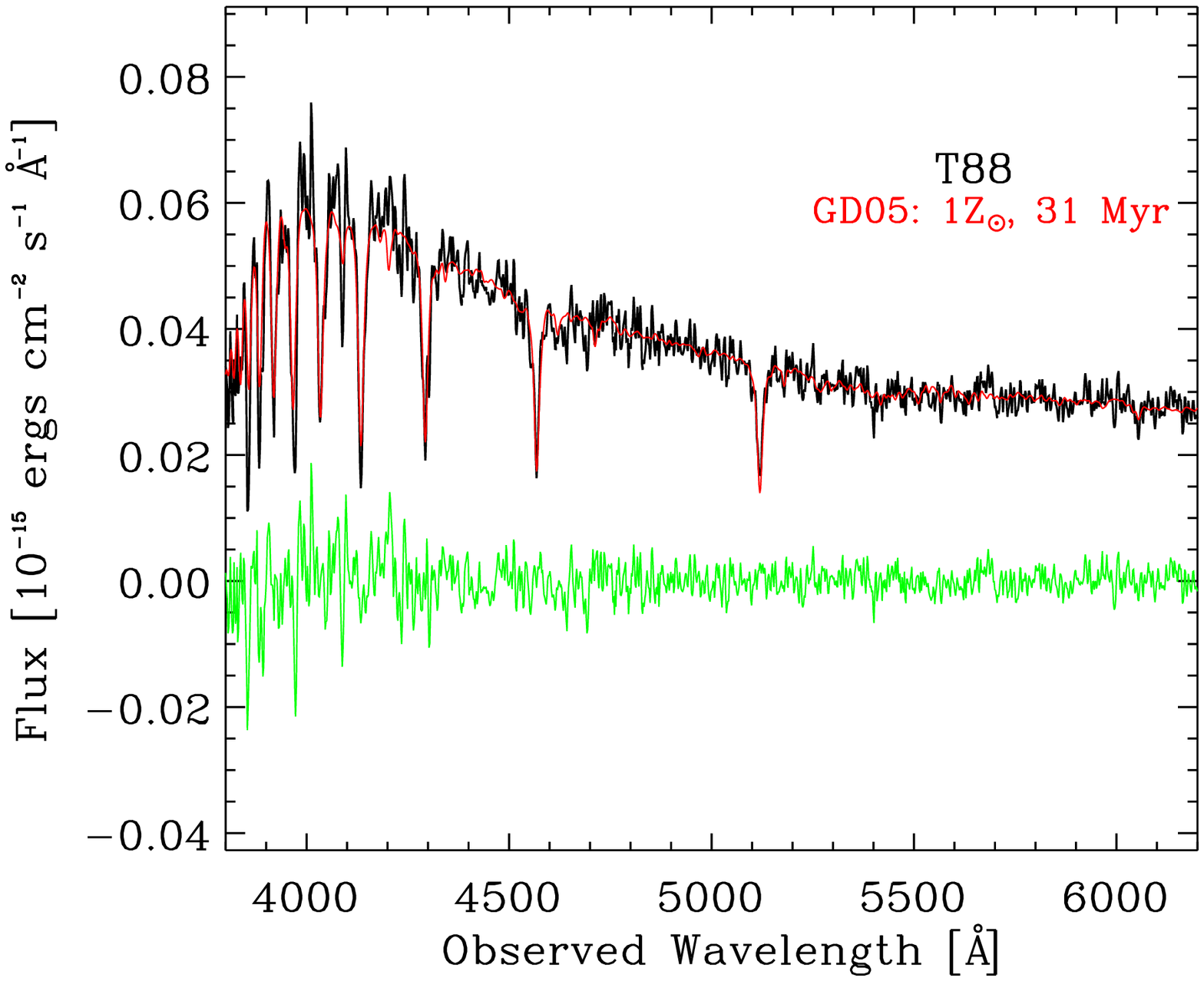}{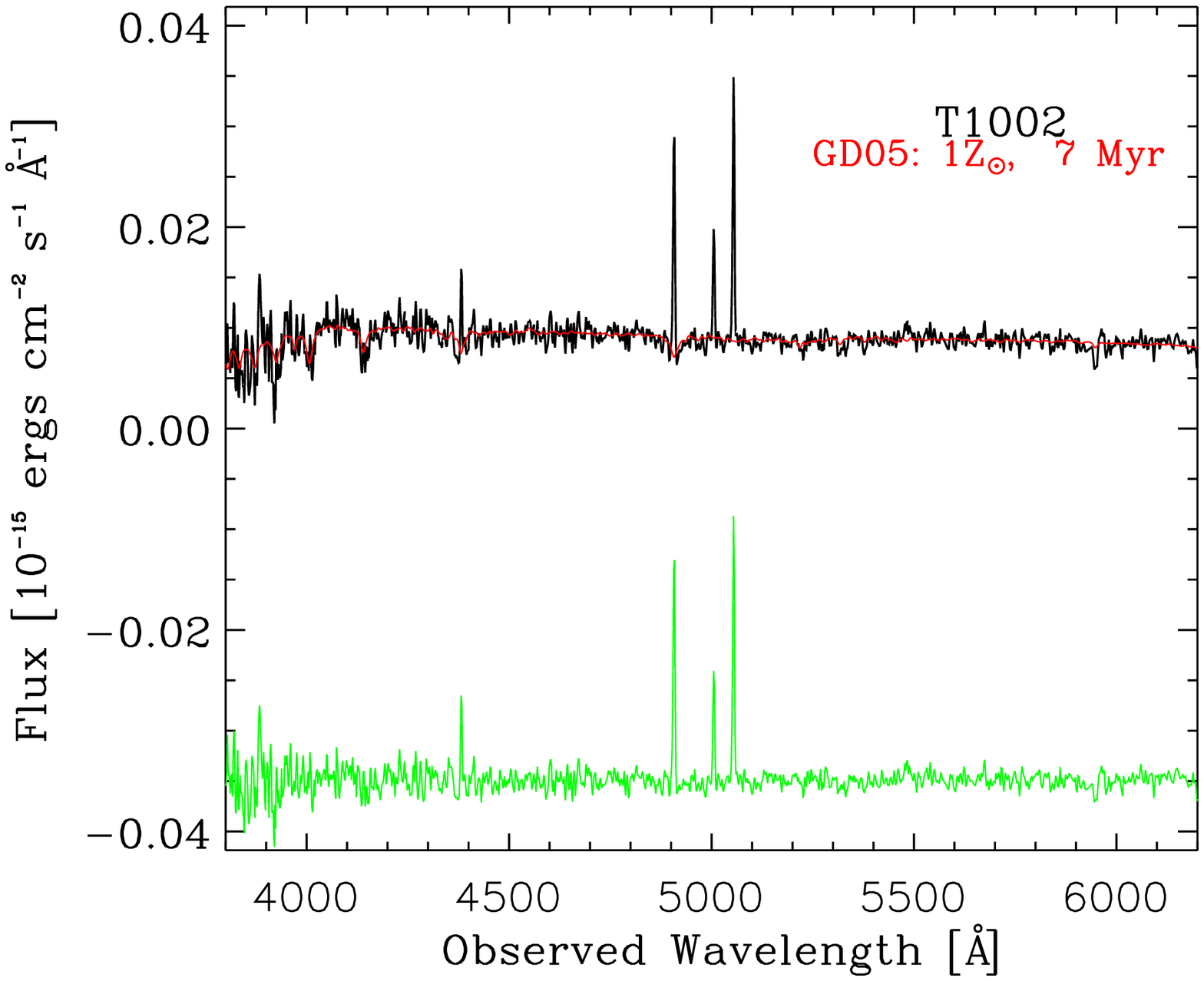}
       \caption{Examples of spectra for two clusters in our sample.
       The observed spectra have been corrected for the estimated
       interstellar extinction.  The red lines represent the best
       fitting (see \S~\ref{sec:ages} for a discussion of the method)
       model template (age and metallicity).  The green lines represent
       the residual (observed cluster $-$ best fitting template $-$ constant).
       The parameters for the best fitting template are given in each panel.}
   \label{fig:examples}
   \end{figure}

\begin{figure}
   \epsscale{0.80}
  \plotone{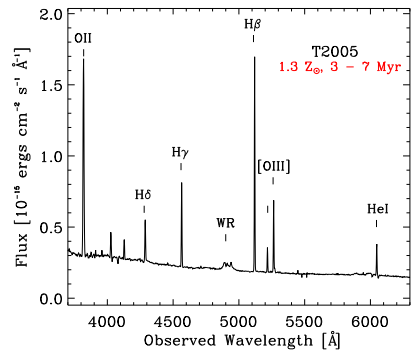}
        \caption{Example spectrum of an emission-line cluster, T2005, which
        also shows strong Wolf-Rayet features.}
   \label{fig:wr}
   \end{figure}

\begin{figure}
 \begin{center}
   \epsscale{1.0}
  \plotone{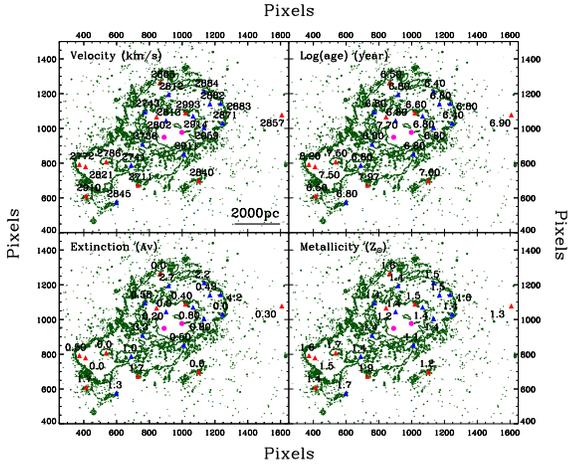}
      \caption{ The position of the clusters in NGC 3256 for different
       velocities (top left),  ages(top right,where the ages are given in logarithmic units in years), extinctions (bottom left),
      and metallicities (bottom right).  The contours are shown to highlight the main features
      of the galaxy.  The upper and lower (magenta) circles mark the nucleus
      and the second brightest optical source in the galaxy, respectively.
      Triangles (red) mark clusters whose spectra are dominated by absorption
      lines while blue marks emission line clusters (see also
      Table~\ref{table:properties1}).} 
    \label{fig:positions}
    \end{center} 
    \end{figure}

\begin{figure}
   \epsscale{1.10}
   \plotone{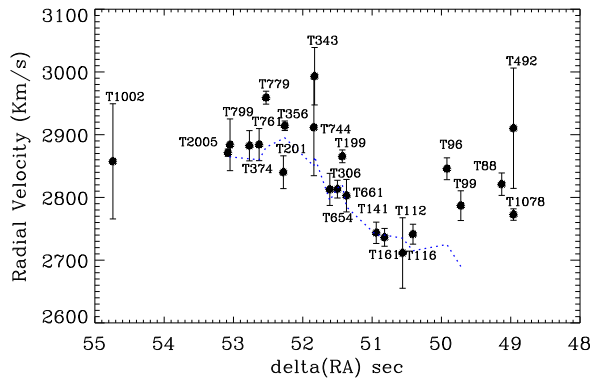}
      \caption{A position--velocity diagram for the observed clusters in
      \n3256.  The dotted line shows the rotation curve of the molecular gas,
       as measured by Sakamoto et al.~(2006).}
   \label{fig:rot-vel}
   \end{figure}
 
\begin{figure}
   \epsscale{1.0}
      \caption{ Metallicity distribution of the clusters in  NGC3256. The figure includes the 
      metallicities derived from both the absorption and emission line clusters.  Note that all the
       clusters are fairly metal rich, with a mean around 1.5 \zo}
   \label{fig:hist}
   \end{figure}
 
\begin{figure}
   \epsscale{1.0}
      \caption{HI position--velocity plot by English et al. (2003), with
the observed clusters superposed.  Open red circles mark the clusters that
kinematically follow the rotating CO disk, while filled blue circles mark those that
do not follow the CO disk, but follow the HI velocities instead.}
   \label{fig:english-himap}
   \end{figure}

\begin{deluxetable}{lcccccccc}

\def\psn{\phs\phn}
\def\pnn{\phn\phn}
\tablecolumns{10}
\tablewidth{-20pt}
\tablecaption{Cluster properties. (The magnitudes have been only corrected for Galactic extinction)}
\tablehead{
\colhead{ID} & \colhead{A/E\tablenotemark{a}} &   \colhead{$\Delta$RA\tablenotemark{b}} & \colhead{$\Delta$DEC\tablenotemark{b}} &  \colhead{$M_{g'}$} & \colhead{$M_{r'}$} & \colhead{A$_V$\tablenotemark{c} }& \colhead{Z}& \colhead{Log(age)}\\
\colhead{ } & \colhead{ } &   \colhead{(sec) } & \colhead{(arcsec) } & \colhead{(mag)} & \colhead{(mag)} & \colhead{(mag)} & \colhead{(\zo) }& \colhead{(year)}
}

\startdata
T88     &0   & 49.13  & 22.04      &-12.8$\pm$0.1  &-12.6$\pm$0.1  &0.00    &1.5$\pm$0.5         &7.5$\pm$0.1    \\ 
T96     &1   & 49.92  & 32.90      &-13.1$\pm$0.1  &-12.4$\pm$0.1  &1.30    &1.7$\pm$0.2         &$<$6.8         \\ 
T99     &0   & 49.72  & 21.03      &-14.8$\pm$0.1  &-14.5$\pm$0.1  &0.00    &1.7$\pm$0.4     &7.5$\pm$0.1     \\ 
T112    &0   & 50.56  & 28.61      &-14.0$\pm$0.1  &-13.1$\pm$0.1  &1.70    &1.9$\pm$0.1         &7.96$\pm$0.08  \\  
T116\tablenotemark{e}     &1   & 50.41  & 22.51      &-15.4$\pm$0.1  &-14.7$\pm$0.1  &1.09    &1.4$\pm$0.2         &$<$6.8         \\
T141    &1   & 50.82  & 17.22      &-14.6$\pm$0.1  &-13.7$\pm$0.1  &3.31    &1.4$\pm$0.2         &$<$6.8         \\
T161    &1   & 50.94  & 8.01       &-15.1$\pm$0.1  &-14.2$\pm$0.1  &0.58    &1.4$\pm$0.2         &$<$6.8         \\ 
T199    &0   & 51.43  & -4.44      &-14.3$\pm$0.1  &-14.1$\pm$0.1  &0.00    &1.6$\pm$0.3          &6.5$\pm$0.1      \\  
T201    &0   & 52.28  & 29.03      &-12.5$\pm$0.1  &-12.4$\pm$0.1  &0.00    &1.2$\pm$0.2          &7.0$\pm$0.1      \\  
T306    &1   & 51.50  & 10.62      &-15.8$\pm$0.1  &-15.3$\pm$0.1  &0.00    &1.4$\pm$0.2         &$<$6.8         \\
T343    &0   & 51.83  & 10.91      &-16.1$\pm$0.1  &-16.0$\pm$0.1  &0.40    &1.5$\pm$0.5         &6.6$\pm$0.1    \\
T356    &1   & 52.26  &10.38       &-15.2$\pm$0.1  &-14.9$\pm$0.1  &0.80    &1.4$\pm$0.2         &$<$6.8         \\
T374    &1   &  52.77  & 7.60      &-12.8$\pm$0.1  &-12.5$\pm$0.1  &0.43    &1.5$\pm$0.2          &$<$6.8        \\ 
T492    &0   & 48.96  & 30.92      &-12.4$\pm$0.1  &-11.3$\pm$0.1  &1.70    &1.4$\pm$0.2     &6.5$\pm$0.1    \\
T654    &1   &  51.61  & 3.67      &-12.8$\pm$0.1  &-12.3$\pm$0.1  &2.70    &1.4$\pm$0.2         &$<$6.8         \\
T661    &0   & 51.37  & 9.22       &-15.2$\pm$0.1  &-14.8$\pm$0.1  &0.20    &1.2$\pm$0.2          &7.7$\pm$0.1     \\
T744    &0   &  51.84  & 21.17     &-12.4$\pm$0.1  &-12.0$\pm$0.1  &0.50    &1.1$\pm$0.2     &6.8$\pm$0.1 \\
T761    &2   &  52.63  & 3.53      &-14.2$\pm$0.1  &-13.6$\pm$0.1  &2.20    &1.5$\pm$0.2         &5.9$-$6.7 \tablenotemark{d}        \\
T779    &1   & 52.53  & 14.00      &-12.9$\pm$0.1  &-12.6$\pm$0.1  &0.80    &1.4$\pm$0.2         &$<$6.8         \\
T799    &1   & 53.05  &  7.78      &-13.0$\pm$0.1  &-12.9$\pm$0.1  & 4.20   & 1.6$\pm$0.2      &$<$6.8       \\
T1002 &0   & 54.74  & 12.46      &-10.6$\pm$0.1  &-10.2$\pm$0.1  &0.30     &1.3$\pm$0.1        &6.9$\pm$0.1    \\
T1078 &0   & 48.96  & 21.17      &-12.9$\pm$0.1  &-12.7$\pm$0.1  &0.50    &1.6$\pm$0.3          &8.2$\pm$0.1 \\  
T2005 \tablenotemark{e}  &2   & 53.08  & 13.20      &-15.4$\pm$0.1  &-14.8$\pm$0.1  &0.00    &1.4$\pm$0.2     &5.9$-$6.7 \tablenotemark{d}        \\

\enddata
\tablenotetext{a}{\,0=absorption, 1=emission, 2=emission with WR features}
\tablenotetext{b}{\,From Base position RA=10:27:00 DEC=$-$43:54:00 (J2000)}
\tablenotetext{c}{\,These extinction have been calculated on spectra already corrected by Galactic extinction ($A_V=0.403$) }
\tablenotetext{d}{\,Ages calculated using the Starburst 99 models}

\label{table:properties1}

\end{deluxetable}

\begin{deluxetable}{lccccccc}
\def\psn{\phs\phn}
\def\pnn{\phn\phn}
\tablecolumns{8}
\tablewidth{0pt}
\tablecaption{Measured indices for the absorption line clusters.}
\tablehead{
\colhead{ID} &\colhead{ $H+He$\tablenotemark{a} } &\colhead{ $K$\tablenotemark{a} } &\colhead{ $H8 $\tablenotemark{a} }  &\colhead{$H\gamma_A$\tablenotemark{b}}    &\colhead{ $Mgb5177$\tablenotemark{b}}    &\colhead{ $Fe5270$\tablenotemark{b}}    &\colhead{ $Fe5335$\tablenotemark{b}} \\
\colhead{} & \colhead{ (\AA) } & \colhead{ (\AA) } & \colhead{ (\AA) }  & \colhead{( \AA)}& \colhead{( \AA)}& \colhead{( \AA)}& \colhead{( \AA)} 
}
\startdata
T88      &6.45$\pm$0.31  &0.34$\pm$0.25   &5.75$\pm$0.31    &6.26$\pm$0.31       &0.22$\pm$0.17  &0.84$\pm$0.22   &1.48$\pm$0.28     \\  
T99      &6.40$\pm$0.44  &0.35$\pm$0.26   &5.64$\pm$0.43    &6.25$\pm$0.28       &0.25$\pm$0.17  &0.88$\pm$0.20   &1.59$\pm$0.28     \\     
T112     &8.53$\pm$0.40  &0.79$\pm$0.23   &7.52$\pm$0.42    &7.88$\pm$0.22       &0.41$\pm$0.10  &1.18$\pm$0.12   &1.81$\pm$0.18     \\     
T199     &3.09$\pm$0.31  &0.03$\pm$0.17   &2.40$\pm$0.31    &3.18$\pm$0.21       &0.16$\pm$0.12  &0.50$\pm$0.16   &0.99$\pm$0.23     \\     
T201     &4.21$\pm$1.51  &0.11$\pm$0.85   &3.39$\pm$1.49    &4.19$\pm$1.01       &0.45$\pm$0.11  &0.81$\pm$0.26   &1.08$\pm$1.09     \\     
T343     &2.96$\pm$0.44  &0.40$\pm$0.24   &2.17$\pm$0.41    &3.33$\pm$0.33       &0.43$\pm$0.18  &0.61$\pm$0.23   &0.77$\pm$0.34     \\     
T492     &2.31$\pm$0.21  &0.57$\pm$0.21   &2.62$\pm$0.41    &2.87$\pm$0.34       &0.35$\pm$0.08  &0.54$\pm$0.10   &0.71$\pm$0.15     \\     
T661     &7.29$\pm$0.49  &0.63$\pm$0.29   &6.44$\pm$0.48    &7.00$\pm$0.36       &0.39$\pm$0.21  &0.96$\pm$0.26   &1.40$\pm$0.38     \\      
T744     &4.24$\pm$0.46  &0.52$\pm$0.24   &3.10$\pm$0.52    &3.27$\pm$0.30       &0.41$\pm$0.18  &0.75$\pm$0.22   &0.92$\pm$0.33     \\     
T1002    &4.26$\pm$1.60  &0.87$\pm$0.23   &3.55$\pm$0.73    &3.44$\pm$0.47       &0.36$\pm$0.03  &0.65$\pm$0.14   &0.79$\pm$0.17    \\  
T1078    &9.16$\pm$0.33  &0.94$\pm$0.19   &8.23$\pm$0.33    &8.55$\pm$0.22       &0.53$\pm$0.14  &1.20$\pm$0.17   &1.66$\pm$0.26    \\

\enddata
\tablenotetext{a}{\,Index definition by Schweizer \& Seitzer (1998).}
\tablenotetext{b}{\,Lick index.}
\label{table:indices}
\end{deluxetable}


\begin{deluxetable}{lccccc}
\tablecaption{Kinematics and Masses of the clusters.}
\tablewidth{-10pt}

\tablehead{
\colhead{} & \colhead{}  & \colhead{$cz$(CO)} & \colhead{$cz_{\rm hel}$} & \colhead{$\Delta cz$}  &  \colhead{Mass}  \\ 
\colhead{ID} & \colhead{D\tablenotemark{a}} & \colhead{(km/s)} & \colhead{(km/s)} & \colhead{(km/s)} & \colhead{10$^{5}$\msun} } 

\startdata
T88     & 1  & ...          & 2821.1$\pm$17.9	 &...                     & 8.3$\pm$2.6\\
T96     & 1  & 2725:  & 2845.7$\pm$17.4	 &+120:              & 1.7$\pm$0.9\\
T99\tablenotemark{b}     & 1  & 2660   & 2786.9$\pm$23.7	 &+126              &49.5$\pm$1.8\\
T112\tablenotemark{b}   & 0  & 2735   & 2711.4$\pm$56.2	 &-23	     &45.0$\pm$2.4\\
T116\tablenotemark{b}   & 0  & 2715   & 2741.4$\pm$15.9	 &+26	                 &13.8$\pm$7.3\\
T141   & 0  & 2740   & 2736.4$\pm$14.2	 &-4   	                   & 6.7$\pm$3.5\\
T161   & 0  & 2740   & 2743.6$\pm$17.0	 &+3	                             &11.0$\pm$5.8\\
T199   & 1  & 2820:  & 2865.5$\pm$10.2	 &+45:                & 3.4$\pm$0.3\\
T201   & 1  &...          & 2840.2$\pm$26.1	 &...                     & 2.3$\pm$0.7\\
T306\tablenotemark{b}   & 0  & 2815   & 2813.2$\pm$14.1	 &-2	                       &19.9$\pm$0.1\\
T343   & 1  & 2865   & 2993.2$\pm$45.8	 &+128	        &18.6$\pm$1.6\\         
T356\tablenotemark{b}   & 0  & 2895   & 2914.2$\pm$7.6	 &+19	               &12.1$\pm$6.4\\
T374   & 0  & ...          & 2882.4$\pm$24.1	 &...                            & 1.2$\pm$0.6\\
T492   & 1  & ...          & 2910.3$\pm$95.9	 &...               & 2.2$\pm$0.0\\
T654   & 0  & 2795   & 2812.8$\pm$25.6	 &+22	                  & 1.3$\pm$0.7\\
T661\tablenotemark{b}  & 0  & 2785   & 2802.8$\pm$25.9	 &+17	          &35.5$\pm$8.3\\
T744   & 0  & 2850   & 2911.7$\pm$77.1	 &+61	     &2.1$\pm$0.5\\
T761   & 0  & 2860   & 2884.2$\pm$25.6	 &+25	               &4.6$\pm$2.4\\
T779   & 1  & 2880   & 2959.0$\pm$10.3	 &+79	                   &1.4$\pm$0.7\\
T799   & 0  & ...         & 2883.9$\pm$41.2	 &...                             &1.6$\pm$0.8\\
T1002 & 0  & ...         & 2857.5$\pm$91.8	 &...                      &0.2$\pm$0.1\\
T1078 & 1  & ...         & 2772.7$\pm$9.1	 &...                 &15.9$\pm$0.6\\
T2005\tablenotemark{b} & 0  & 2865   & 2871.3$\pm$5.8	 & +4                    &14.0$\pm$7.4\\

\enddata
\tablenotetext{a}{\,0=disk, 1=not disk}
\tablenotetext{b}{\,Complexes}

\label{table:properties2}
\end{deluxetable}


\clearpage
\appendix

\section{Appendix A: Reduction of GMOS Data}

The data were reduced using the Gemini {\sc IRAF} package.
The individual tasks used at each of the reduction steps are listed below in parentheses.
The raw GMOS images (3 CCDs read out using one amplifier each)  are multi-extension 
FITS (MEF) files with one primary
header unit, containing all the usual header information, and three pixel
extensions, one for each of the detectors.

\subsection{Imaging Data}

Mean bias frames were created combining all the available bias frames
taken during each of the GMOS-S observing runs (task {\tt gbias}). 
   
Twilight flat fields were created from all the available twilight frames
taken during each of the GMOS-S observing runs ({\tt giflat}).
The overscan section of the images was trimmed off, and the images were
bias-subtracted and flat-fielded using the task {\tt gireduce}.
The three image extensions in each exposure were then mosaiced into a
single image extension in which the shifts and rotations between the
three CCDs have been removed ({\tt gmosaic}).
     
The images were registered and co-added using the task {\tt imcoadd}, and they
have been calibrated using the science exposures taken under photometric
conditions.

\subsection{ Spectroscopic Data}

We are using the MOS mode with approximately 25 spectra
per mask.  Each set of  data include an arc
and flat taken along with each science exposure and spectral dithers.
Bryan Miller developed our own script-driven pipeline {\sc MOSPROC}, based on Gemini
{\sc IRAF} scripts and customs IDL routines.
The  Gemini package scripts have been modified,
mainly to improve the propagation of data quality planes.

\begin{itemize}

\item{ MOSPROC}

\begin{itemize}
\item{ Bias substraction for spectra, flats and arcs }

The bias subtraction for the spectroscopic data was done in the same way
as for the imaging.
Quartz-halogen flat fields exposures were taken during the night, either
after or before the science exposures. 
To run through {\sc mosproc}, the science, arcs, and flats frames
were first bias subtracted using the task {\tt gsreduce}.

\item{ Bad Pixel Mask}

A bad pixel mask was constructed from the quartz-halogen flats, taking
into account the known bad pixels in each CCD, and masking the existing
emission lines in the spectral QH lamps that came from
the IR diffuser.

\item{ Flat-Field Correction}
   
The quartz-halogen flats were overscan trimmed and bias-subtracted
like the imaging data and the mask definition file (MDF) was added as
a table extension to the MEF.  The MEF contains the locations of the
slits in the focal plane and is used for bookkeeping during the
remaining reduction.  The flats were normalized by fitting a high
order polynomial to each line to remove the shape of the quartz lamp
but leaving the fringe pattern at the red end of the spectra
 {\tt gsflat}. The task {\tt gsreduce} was then used to divide the science data
by the flat fields.
   
The resulting science exposures still have three extensions, but now
trimmed, bias-subtracted, and flat-fielded.

\item{Wavelength Calibration and Distortion Correction }
 
 The wavelength calibration was determined from CuAr lamp spectra taken
either before or after the science exposures.  The dispersion function
was fit with a sixth-order polynomial that gave a typical rms error of
0.3 Angstroms ({\tt gswavelength}).  Within each slitlet the position of each
arc line with spatial position is used to rectify and wavelength
calibrate each 2D spectrum ({\tt gstransform}).  

 \item{ Cosmic Rays Removal}
 
Cosmic rays were identified and removed using a Laplacian edge
detection algorithm  \citep{2001PASP..113.1420V}.  The locations of the cosmic
rays are recorded in the 2D data quality image for each slitlet.

The spectra were then traced, background-subtracted, and extracted
using {\tt gsextract}, a wrapper for {\tt twodspec.apall}, to allow the handling
of MEF files.  The spectra were traced using a 5 order polynomial
and averaging every 50 pixels in the dispersion direction. Background
subtraction was done by fitting a second-order polynomial with 3$\sigma$\ 
rejection to a region perpendicular to the trace.  The variance and
background spectra were saved to help with the error estimation.

\item{Quantum Efficiency (QE) Correction}

The quantum efficiency (QE) as a function of wavelength for the three GMOS
CCDs can differ by up to about five percent at the given wavelength.
If this difference is not corrected then spectra can have noticable
jumps at the gaps between the CCDs.  We have measured the relative QE
of CCD1 and CCD3 with wavelengh compared with CCD2 using QH flats
taken at 25nm intervals from 350nm to 700nm.  The corrections are
applied to the data using an IRAF script called  {\tt qecorr}.

\end{itemize}

\item{Correction for Slit Losses}

The GMOS instruments do not have atmospheric dispersion correctors and
in general the MOS spectra were not taken with the slits parallel to
the parallactic angle. Therefore, there are wavelength-dependent slit
losses due to the difference between the parallactic angle of each
exposure and the position angle of the slits.  In order to combine the
spectra from different exposures properly these difference must be
corrected.  We calculate the slit losses based on measured image
quality, slit width (0.75 arcsec), and the PA-parallactic angle
difference using the method of   \cite{1982PASP..94.715V}.

\item{ Relative Flux Calibration and Reddening}

\begin{itemize}

\item{Relative Flux Calibration}

The relative instrument spectral response function was determined
using observations of a flux standard star.  Spectra were taken at
central wavelengths of 400, 500, and 600nm so that the combined
sensitivity function covered the full wavelength range of the MOS
spectra.  The standard spectra were reduced in exactly the same way as
the MOS spectra.  The sensitivity function was computed using the IRAF
task gssensfunc and this was applied to the MOS spectra using the task
{\tt gscalibrate}.

As the last step, we combined the individual spectra of the same
sources using our IRAF task called gscombine.  This uses the final
data quality planes to mask the chip gaps and other bad pixels and
then scales and averages the spectra using {\tt scombine}.  
         
\item{Reddening}
 
The interstellar extinction along the line of sight towards \n3256 is
$A_V=0.403$.
The final calibrated spectra were corrected using the empirical selective
extinction function from \cite{1989ApJ...345..245C}   included in the
task {\tt noao.onedspec.deredden}.
    
\end{itemize}
\end{itemize}
\begin{acknowledgements}

G.T. would like to  thank to Matt Mountain and Phil Puxley for the tremendous support throughout this project.

Based on observations obtained at the Gemini Observatory, which is operated by the
Association of Universities for Research in Astronomy, Inc., under a cooperative agreement
with the NSF on behalf of the Gemini partnership: the National Science Foundation (United
States), the Particle Physics and Astronomy Research Council (United Kingdom), the
National Research Council (Canada), CONICYT (Chile), the Australian Research Council
(Australia), CNPq (Brazil) and CONICET (Argentina)
\end{acknowledgements}


\end{document}